\documentstyle[preprint,aps,amssymb,epsf]{revtex}  
\tightenlines

\begin{document}
\draft                           


\title{Twenty years since the discovery of the Fractional Quantum Hall Effect: Current state of the theory}                           

\author{M.I. Dyakonov}  

\address{Laboratoire de Physique Math\'ematique,  Universit\'e 
         Montpellier 2,  France}


\maketitle


\begin{abstract}

The current state of the theory of the Fractional Quantum Hall Effect is critically analyzed, especially the generally accepted concept of composite fermions. It is argued that there is no sound theoretical foundation for this concept. A simple one-dimensional model is proposed, which presumably has an energy spectrum similar to that of the FQHE system.

\end{abstract}

---------------------------------------------------------------

\vskip 2cm

{\bf 1. Introduction}\\

Back in 1982, when condensed matter physicists were slowly recovering from the shock produced by the discovery of the (integer) Quantum Hall Effect (IQHE) two years earlier, \cite{klitzing} a new surprise came in a paper by Tsui, Stormer, and Gossard, \cite{tsui} who reported quantized Hall plateaus at filling factors $\nu=1/3$ and $\nu=2/3$ (see Fig. 1).  This finding opened the vast field of exciting studies of the Fractional Quantum Hall Effect (FQHE), in which many new surprises were to come.

 It was clear from the start that the pronounced features observed at fractional fillings are due to the effects of electron-electron interactions.  In Ref. 2 it was suggested that a new electronic state is formed, "such as a Wigner solid or a charge density wave with a triangular symmetry". However, calculations of the ground state energy of a Wigner crystal in the lowest Landau level by Yoshioka and Lee \cite{yoshioka} did not show anything special at filling $\nu=1/3$.  A strikingly new explanation was put forward in the famous paper by Laughlin, \cite{laughlin} who proposed a $\nu=1/3$ ground state wavefunction minimizing the energy and corresponding to a uniform electron density (the incompressible quantum fluid).  He showed that a gap should exist in the excitation spectrum and introduced fractionally charged quasi-particles which should appear when the filling slightly deviates from the value $\nu=1/3$.  (Quasi-particles with charge $e/3$ were mentioned already in Ref.  \cite{tsui} and were  previously  shown  to exist in 1/3-filled quasi one-dimensional systems by Su and Schrieffer  \cite{su}).  Laughlin predicted other similar states for $\nu=1/(2m+1)$.

While the $\nu=1/5$ and $\nu=1/7$ plateaus were indeed observed later, the reality appeared to be much more complicated and interesting.  Fig. 2 from Stormer's article \cite{stormer} (see also Ref. \cite{du}) presents the picture of the Fractional Quantum Hall Effect as seen today.  Comparing
\begin{figure}

\hskip 50pt \epsfxsize=250pt{\epsffile{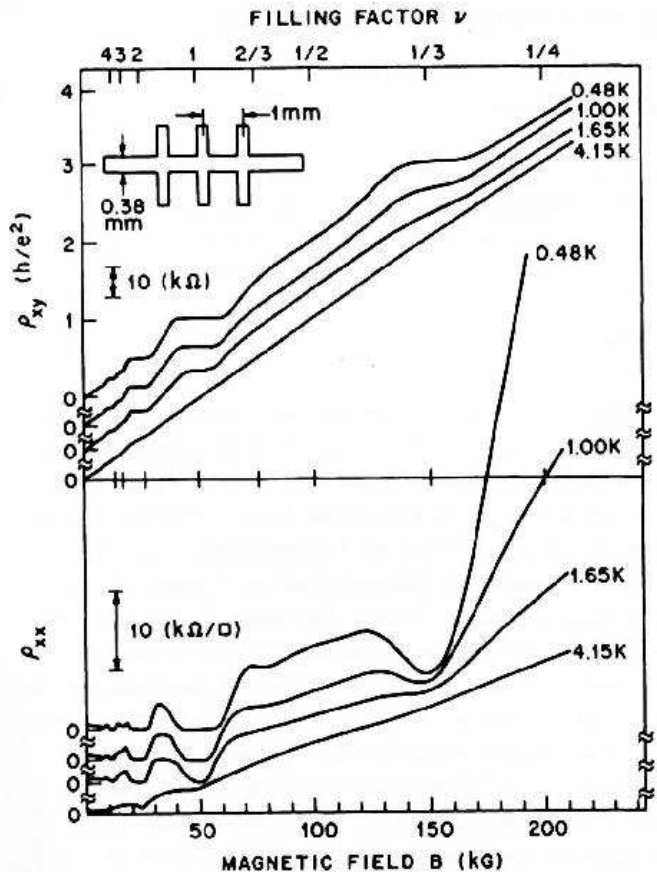}}
 
\vskip 20pt
 
\caption{First observation of the Fractional Quantum Hall Effect (from Ref. 2).}
 
\end{figure}
\noindent Figs. 1 and 2 one can see the amazing progress achieved during 10 years by increasing the mobility from $10^{5}$ to $10^{6}$ cm$^{2}$/V.s and lowering the temperature from 0.5 K to 40 mK, i.e. by enhancing the role of electron-electron Coulomb interaction compared to the random potential and thermal energy. The most pronounced features appear at $\nu=1/3, 2/5, 3/7, 4/9...$ converging to $\nu=1/2$ (currently referred to as Jain sequence). 
                                                                                Encouraged by Laughlin's success in describing the $\nu=1/3$ state, many theorists attempted to guess the good wavefunctions, and various schemes for describing the new states were proposed.  For some time the Haldane hierarchy \cite{haldane} of incompressible fluid states was popular, in which quasi-particles existing in the vicinity of a given Laughlin state condense to a new Laughlin state with electrons replaced by quasi-particles, and this construction may be iterated many times. 

 A major step forward was made by Jain \cite{jain} who, in the spirit of "wavefunction engineering" proposed wavefunctions for the states with $\nu=p/(2mp+1)$ with integer $p$ and $m$.  What is much more important, Jain advanced a novel idea: the FQHE of electrons is a manifestation of the IQHE for "composite fermions".  The interacting electrons can be re-
\begin{figure}

\hskip 15pt \epsfxsize=450pt{\epsffile{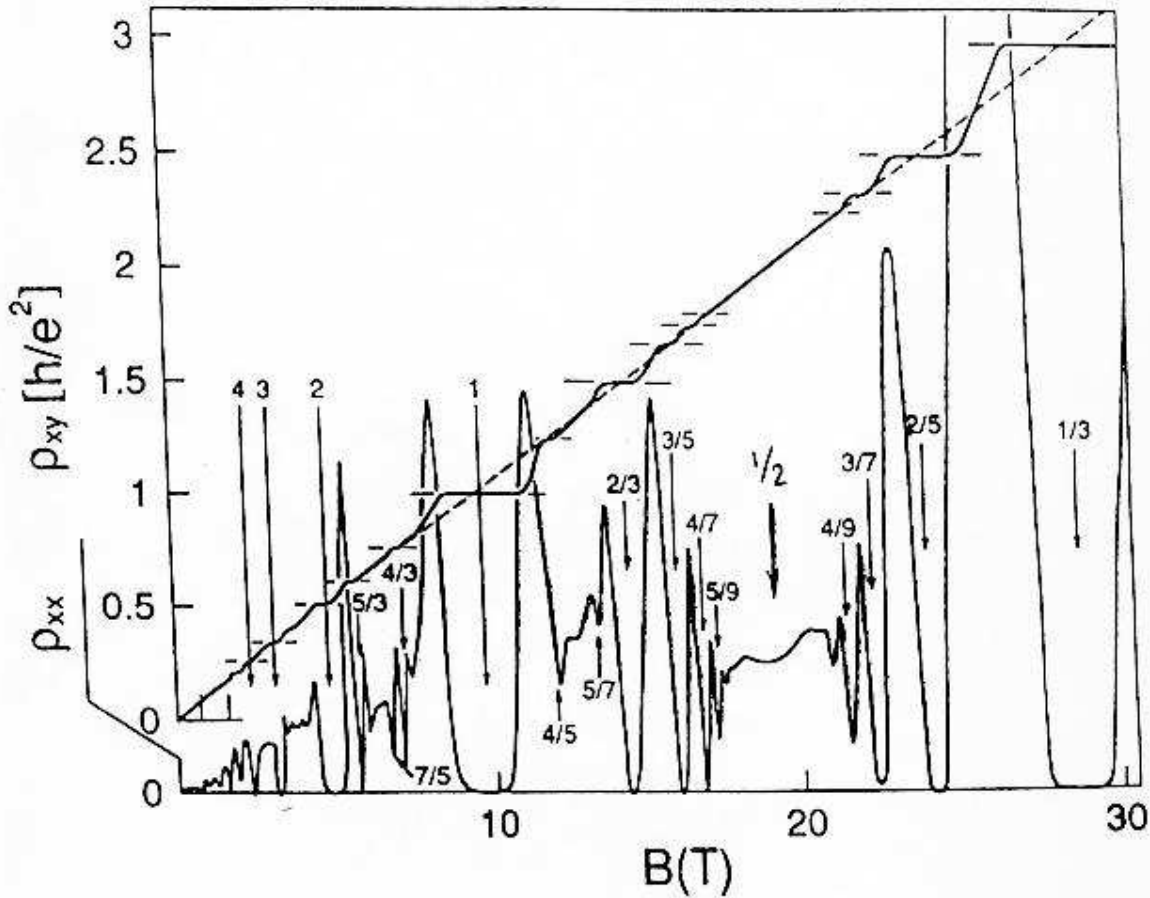}}
 
\vskip 15pt
 
\caption{The developed Fractional Quantum Hall Effect (from Ref. 6)}
 
\end{figure}
\noindent placed by non-interacting (in the first approximation) composite fermions, which see a reduced magnetic field, such that the composite fermion filling factor, $\nu^{\star}$, is related to the electron filling factor by $1/\nu = 1/\nu^{\star}  + 2m$, so that for $m=1$, the values $\nu^{\star}=1, 2, 3...$ correspond to $\nu=1/3, 2/5, 3/7...$ 

This idea was further developed by Lopez and Fradkin,  \cite{lopez} Kalmeyer and Zhang,  \cite{zhang} and in a more detailed way by Halperin, Lee, and Read  \cite{halperin} by introducing a singular gauge transformation replacing electrons by composite fermions, thus formalizing Jain's idea of a reduced effective magnetic field seen by these new quasi-particles.  It was suggested in Refs. \cite{zhang} and \cite{halperin} that the $\nu=1/2$ state, as well as other states at even denominator fillings, are in fact Fermi liquids with zero effective magnetic field. 

The concept of composite fermions moving in a reduced magnetic field is now widely accepted and is the basis of the current understanding of the Fractional Quantum Hall Effect.  The power of this concept can be appreciated by looking at Fig. 2.  The curves for $\rho_{xx}$ and $\rho_{xy}$ for $B>18$ T, are very similar to what is observed in the integer QHE, if we put the zero of the effective field at 18 T, where $\nu=1/2$. For $\nu$ slightly exceeding this value, the system behaves like degenerate electrons in a weak magnetic field, with characteristic Shubnikov-de Haas oscillations, which later develop into the QHE.  Many other striking experimental facts confirm this concept (see the review by Willett \cite{willett}), in particular the cyclotron radius corresponding to the effective magnetic field was observed.

Does this mean that the concept of composite fermions is firmly established, and has a sound theoretical foundation?  I don't think so, and it is the purpose of this presentation to show that in fact nobody has really derived theoretically the existence of composite fermions, nor even explained what is a composite fermion.  Furthermore, there are a number of simple questions, which remain without reasonable answers.  I will argue that after 20 years of intense experimental and theoretical efforts, and something like 2000 papers devoted to the subject, our understanding of the Fractional Quantum Hall Effect is still very limited, although certainly an enormous progress has been made due to the insight and intuition of people cited above, as well as of many others.

The field of FQHE includes many very interesting topics, such as spin phenomena, charge density waves and anisotropic transport in higher Landau levels, bilayer systems, edge transport, etc, which are beyond the scope of this presentation.
 
I will focus on the basic theoretical problem of FQHE: completely spin-polarized interacting electrons, partly filling the lowest Landau level, and I will critically analyze the main theoretical results in this domain (which, of course, is much easier than to make a meaningful contribution to the field).  In order to have the generally accepted point of view on the subject, the reader is invited to consult recent reviews \cite{fermions} - \cite{shankar} devoted to composite fermions. The reader is supposed to be familiar with the phenomenology of the Quantum Hall Effect, as well as with the quantum mechanics of an electron in a magnetic field.\\  
\vskip 0.6cm

{\bf 2. The problem}\\ 

We have $N$ two-dimensional electrons in a plane perpendicular to magnetic field $B$.  The number of states at a given Landau level is $M=(eB/2\pi\hbar c)S$, where $S$ is the sample area, other notations are conventional, and $N<M$.  The magnetic field is assumed to be so strong that the distance between adjacent Landau levels, as well as the spin Zeeman energy, are much greater than the characteristic energy of the Coulomb electron-electron interactions, $e^{2}/r$, where $r \sim (S/N)^{1/2}$ is the mean distance between electrons.   In this case the electrons may be regarded as fully spin-polarized (or spinless) fermions confined to the lowest Landau level.  Mixing with higher Landau levels can be ignored, and they are absolutely irrelevant.  We are interested in finding the energy spectrum and the wavefunctions of this system for arbitrary filling $\nu=N/M <1$. 

Thus we have a strongly interacting many-particle system, and our problem is of enormous difficulty.  After introducing natural units of length, $l=(\hbar c/eB)$ , and energy, $e^{2}/l$, we are left with a dimensionless problem of diagonalizing a huge numerical matrix with a single dimensionless parameter, $\nu$.  There are also no visible symmetries in the problem, except trivial ones, like the particle permutation symmetry, which requires that the many-particle wavefunction be antisymmetric, and the conservation of the total angular momentum (in the disk geometry).  If $\nu$ were very small, one could use it as a small parameter, and indeed there is theoretical, as well as experimental, evidence that for small enough  $\nu$ a Wigner crystal is formed.  However the most interesting phenomena occur when $\nu$ is {\it not} very small.   In this case our purely numerical problem appears intractable theoretically, since no approximation can be justified, and we have to rely on numerical calculations for small values of $N$ and $M$ (the difficulty of such calculations increases exponentially, and become unsurpassable for $N$ greater than 10-15).

It would be only of minor interest to know the exact numerical values of the ground state and first excited state energies (in units of $e^{2}/l$), if we were not aware of the totally unexpected {\it experimental} fact that something special happens at rational values of $\nu$, namely, that gaps in the excitation spectrum appear when $\nu=p/q$ with $q$ odd.  (The $N=10^{9}$ electrons in a Hall bar at 40 mK diagonalize their Hamiltonian very quickly and show us the final result).  There should exist a general way of understanding this exact result for the eigenvalues of our numerical matrix, and the composite fermion concept is designed to provide such understanding.  In what follows we will see that this goal is not achieved so far.
 
It is common to use the disk geometry, the symmetrical gauge, and the following basis of one-electron states in the lowest Landau level: 
 $$\phi_{k}(z)=(2^{k+1}\pi k!)^{-1/2}z^{k}\exp(-|z|^{2}/4), \eqno{(1)}$$
with $z=x+iy$, $x$ and $y$ being the electron (dimensionless) coordinates,  $k= 0, 1, 2 ...M-1$ labels the $M$ states of one electron belonging to a disk of radius $\sqrt{2M}$. Hence an arbitrary many-particle state in the lowest Landau level is described by a wavefunction
$$\Psi(z_{1},... z_{N})=P(z_{1},... z_{N})\exp(-\frac{1}{4} \sum|z_{i}|^{2}), \eqno{(2)}$$
where $P$ is an antisymmetric polynomial in the electron coordinates $z_{1}... z_{N}$ of a maximal degree in each variable, equal to $M-1$. To ensure a well-defined total angular momentum (which is a conserved property), $P$ must be homogeneous.  For a completely filled lowest Landau level ($\nu=1$), there is a single possibility: $P(z_{1},... z_{N})={\prod_{i>j}} (z_{i}-z_{j})$ .  Thus, the general problem is to find these polynomials, as well as the corresponding energies, for any state at arbitrary $\nu$; in particular, for the ground state $P$ should be chosen to minimize the interaction energy $V={\sum_{i<j}} U(z_{i}-z_{j})$, where $U$ is the potential energy of a pair of electrons.\\

\vskip 0.6cm

{\bf 3. The Laughlin states}\\

Probably the most well established theoretical idea in the domain of the FQHE is the Laughlin wavefunction \cite{laughlin}  describing the ground state at $\nu=1/(2m+1)$:  
		
$$\Psi_{1\over 2m+1}(z_{1},... z_{N})={\prod_{i<j}} (z_{i}-z_{j})^{2m+1}\exp(-\frac{1}{4} \sum|z_{i}|^{2}). \eqno{(3)}$$

This (un-normalized) wavefunction is antisymmetric, and by counting the maximal power of each $z_{i}$, which is $(2m+1)(N-1)$, one verifies that $\nu=N/M$ tends to $1/(2m+1)$, as $N\rightarrow \infty$.  The advantage of the Laughlin function is that for $m=1,2...$ it goes to zero at $z_{i}\rightarrow z_{j}$ as $(z_{i}-z_{j})^{2m+1}$, faster than an arbitrary antisymmetric function, thus minimizing the electron repulsion and the ground state energy. \cite{note1}
  
Thus Laughlin gave a simple and clear answer to the fundamental question: {\it why does something special happen at $\nu=1/3$?} - Because, for this filling a ground state wavefunction may be constructed, which goes to zero at $z_{i} \rightarrow z_{j}$ faster than for neighboring values of $\nu$.  This circumstance is responsible for a gap in the energy spectrum.
  
Laughlin has also introduced fractionally charged quasi-particles, which appear if, starting (for example) at $\nu=1/3$ and leaving $N$ unchanged, we increase or reduce the number of states, $M$, by one (quasi-holes or quasi-electrons, respectively).  Laughlin's wavefunction for a quasi-hole located at $z_{0}$ is given by:

$$\Psi_{1/3}^{(z_{0})}=A_{z_{0}}\Psi_{1/3}, \hspace{0.2in} A_{z_{0}}={\prod_{i}}(z_{i}-z_{0}). \eqno{(4)}$$
The energy of the quasihole should not depend on $z_{0}$.

While these ideas certainly gave a clue to understanding the FQHE, some important questions remain unanswered.  One of them concerns the $\nu=2/3$ state (and, generally, $\nu=1-1/(2m+1)$ states).  Because of the electron-hole symmetry, this state can be regarded as the $\nu_{h}=1/3$ hole state, so the wavefunction can be written in the form of Eq. (3), where now $z_{i}$ should be considered as the coordinates of $N$ holes in a completely filled Landau level.  The physical properties should be (and, in fact, are) quite similar to those at $\nu=1/3$.  There is a simple relation between the correlation functions and the ground state energies at fillings $\nu$ and $1-\nu$. \cite{note2}

Suppose, however, that one wants to have a look at this  function written in terms of $2N$ {\it electron} coordinates.  To do this, one must (i) write down the hole function in Eq. (3) as a superposition of $N\times N$ determinants involving one-particle hole wavefunctions given by Eq. (1), and (ii) leaving the coefficients in the superposition unchanged, replace each determinant by its complimentary $2N \times 2N$ electron determinant.  The resulting unwieldy expression, which nobody knows how to write down explicitly, will represent the $\nu=2/3$ ground state, $\Psi_{2/3}$. It will not have any elegant form comparable to Eq. (3), it will go to zero at  $z_{i} \rightarrow z_{j}$ as  $(z_{i}-z_{j})$, just like any antisymmetric function, and we will hardly be able to understand why this function should minimize the interaction energy!  This shows the existence of wavefunctions that are as good as the Laughlin function, but which do not have higher order zeros when the electron coordinates coincide.  In this sense $\Psi_{2/3}$ resembles the wavefunctions for other odd-denominator fillings, such as $\Psi_{2/5}$. Nobody knows what are the relevant properties of these ground state wavefunctions. 

Another question concerns Laughlin quasiparticles. The parameter  $z_{0}$ in Eq. (4) is arbitrary, and functions with different values of $z_{0}$ are not orthogonal.  To obtain an orthogonal set one can expand $A_{z_{0}}$ in terms of $z_{0}$ \cite{laughlin}:
$$A_{z_{0}}={\sum_{n=0}^{N}}(-1)^{N-n}A_{n}(z_{1},...z_{N}) z_{0}^{N-n}, $$
where $A_{n}$ are symmetric polynomials of degree $n$ ( $A_{0}=1,  A_{1}=z_{1}+...+z_{N},... A_{N}= z_{1}z_{2}...z_{N}$). Then the functions $\Psi_{1/3}^{(n)}=A_{n}\Psi_{1/3}$ with $n=1, 2, ...N$ provide the basic orthogonal set of $N$ (not $M$) degenerate quasihole states. They are orthogonal because these states correspond to different total angular momenta.  Since the total angular momentum is a conserved quantity, degeneracy of states with different angular momenta normally would be a consequence of some additional symmetry. However, the nature of this symmetry is unknown (a similar question arises within the composite fermion picture, see Section 5).

Thus, even for the Laughlin states, there are some simple questions without any answers,  and this is a clear signal that our understanding is not complete.\\

\vskip 0.6cm

{\bf 4. Jain states and "composite fermions"}\\

Using the analogy with Laughlin function, which may be written as
$$\Psi_{1\over 2m+1}(z_{1},... z_{N})={\prod_{i<j}} (z_{i}-z_{j})^{2m}\Psi_{1}, \eqno{(5)}$$
where $\Psi_{1}$ is the wavefunction of $N$ electrons completely filling the lowest Landau level, Jain \cite{jain} suggested, as a generalization, the following ground state wavefunctions for $\nu=p/(2mp+1)$:

$$\Psi_{p\over 2mp+1}(z_{1},... z_{N})=P_{LLL}{\prod_{i<j}} (z_{i}-z_{j})^{2m}\Psi_{p}, \eqno{(6)}$$                       .			   	

Here $\Psi_{p}$ is the wavefunction of $N$ electrons completely filling $p$ first Landau levels, and $P_{LLL}$ is an operator projecting into the lowest Landau level. The value of $\nu$ can be verified by counting the maximal power of $z_{i}$, which is now equal to $2m(N-1)+N/p$.  Comparing this function with the results of exact diagonalization for several values of $m$ and $p$, Jain and others found a very good overlap, \cite{note3} that justifies the bizarre construction of Eq. (6) involving higher Landau levels, which are physically absolutely irrelevant.  The filling factor, $\nu$, may be written in the form:
                                          				
$${1 \over \nu}=2m+{1 \over p}. \eqno{(7)}$$

On the basis of Eqs. (6), (7) Jain advanced the concept of composite fermions: the FQHE of electrons is a manifestation of the IQHE of composite fermions, the integer $p\equiv \nu^{\star}$ is the filling factor for composite fermions, corresponding to the electron filling factor $\nu$. The value $2m$ ($m=1, 2, 3...$) is the number of magnetic flux quanta "attached" or "bound" to each electron, and different values of $m$ correspond to different sequences of FQHE states.  For example, at $m=1$ one obtains the most prominent sequence $\nu=1/3, 2/5, 3/7... $ Eq. (7) may be rewritten in terms of an effective magnetic field, $B^{\star}$, seen by composite fermions:

$$B^{\star} = B(1- 2m\nu) = B - 2mn\Phi_{0}, \eqno{(8)}$$		      	which, indeed,  looks as if each electron has picked up and neutralized $2m$ elementary magnetic fluxes, $\Phi_{0}$, thus reducing the magnetic field from $B$ to $B^{\star}$.

Interestingly, Eqs. (6), (7) were the {\it only} source of this beautiful picture. Although experiments show that this idea is very productive and certainly corresponds to some reality, one has to state that it is no more (but also no less) than an amazingly successful guess.  Indeed, from the fact, that the construction in Eq. (6), involving $\Psi_{p}$ and projection, provides a good ground state wavefunction, it does not follow logically that some quasi-particles exist, which occupy $p$ Landau levels.  One must have an extraordinary imagination to observe composite fermions completely filling their lowest Landau level, by simply looking at the Laughlin function, (which is a particular case of Jain functions with $m=1, p=1$).
  
This brings us to the basic question, what is a composite fermion?  Jain and Kamilla give the following definition: \cite{kamilla}\\

{\it Composite fermions are electrons carrying an even number of vortices of the many-body wavefunction.}\\
\vskip 0.3cm 
Note that "vortices" is synonym for "zeros", which can be observed in Eq. (6) before projection, but which are destroyed by the $P_{LLL}$ procedure.   A milder definition is: \cite{jain}\\ 

{\it A composite fermion is an electron bound to an even number of flux quanta.}\\ 
\vskip 0.3cm
Finally, I quote the recent review of Simon: \cite{simon}\\

{\it The field of composite fermion physics began in 1989 with a paper by Jainendra Jain who pointed out that there is a mapping between the wavefunctions of integer quantized Hall states and approximate - but extremely good - wavefunctions for the fractional quantized states. This wavefunction mapping can be thought of as binding an even number of vortices (zeros) to each electron, turning it into a "composite" fermion.}\\

The "mapping" referred to in this quotation is given by Eq. (6).  These explanations are not particularly clear: an electron carrying an even number of zeros of the many-body wavefunction is something like an electron carrying its own Hamiltonian.  Equally, it does not seem acceptable for an electron to carry, or bind, some flux quanta, which are not entities that can walk by themselves, and, getting "attached" to lectrons, disappear from the external applied field.  Obviously, this is an attempt to describe some vague thoughts induced by the form of Eq. (6).  It explains neither the nature of the object, whose cyclotron radius (corresponding to the reduced field $B^{\star}$) is seen in experiments, nor even the existence of such objects.\\   

\vskip 0.6cm

{\bf 5. "Chern-Simons" composite fermions}\\

A formal transformation from electrons to composite fermions (singular gauge transformation) was used in Refs. \cite{lopez} - \cite{halperin} to provide a justification for the main statement:  {\it The interacting electrons in external field $B$ can be replaced by non-interacting composite fermions  in effective field $B^{\star}$.}  The essence of this theory may be described in simple words.\\					

a) However large may be the ratio of Landau spacing to the Coulomb energy, we do not restrict ourselves to considering only the lowest Landau level.\\

b) The many-body electron wavefunction is written as the many-body composite fermion wavefunction times a phase factor, which acquires a phase $4\pi m{\cal N}$ when one electron is carried along a loop of area $s$, containing ${\cal N}$ other electrons.\\ 

c) We write the resulting value of this factor as
$$\exp(-4\pi i m{\cal N}) = \exp (-4\pi im({\cal N}/s) s). $$
So far, so good.  However we now observe that ${\cal N}/s \approx n$, where $n$ is the electron concentration.  In the "mean field approximation" we replace:
$$\exp(-4\pi im{\cal N}) \approx \exp(-4\pi imns). \eqno{(9)}$$				

The phase factor in the rhs of Eq. (9) corresponds to the existence of an additional magnetic field. Comparing it with the phase factor which one would obtain by moving a free electron along a loop of area $s$ in a magnetic field $B_{0}$:
$$\exp (2\pi i \frac {B_{0}s}{\Phi_{0}}),$$
we find $B_{0}  = - 2mn\Phi_{0}$, thus the total field is $B^{\star} = B + B_{0}  = B - 2mn\Phi_{0} = B(1-2m\nu)$, as given by Eq. (8).\\ 

d) We drop altogether the Coulomb interaction, assuming that it is mostly taken care of by our transformation.  Thus, in the mean field approximation we have non-interacting composite fermions in the reduced field $B^{\star}$.  The IQHE of composite fermions will occur whenever an integer number of Landau levels in {\it this} field are fully occupied.  For $\nu=1/2m$, we have a Fermi liquid of composite fermions in zero magnetic field.\\

e) Effects of quantum fluctuations around the mean field and effects of the Coulomb interaction presumably largely cancel each other and can be treated by perturbation theory.  These effects may strongly renormalize the properties of composite fermions, such as their mass, but they do not change the overall picture.\\

The reader may check that my "derivation" is equivalent to the original one.  Together with Jain's work, the results  of  Refs. \cite{lopez} - \cite{halperin}, especially the analysis of the $\nu=1/2$ state and its vicinity by Halperin, Lee, and Read, \cite{halperin} gave an extremely simple and appealing picture of the FQHE. By using dimensionality arguments and intuition, Halperin, Lee, and Read succeeded in predicting correctly the sequence of gaps for the FQHE states, and, most impressively, they obtained the $\rho_{xx} \sim q$ relation for the diagonal resistivity at large wave vectors $q$, which explains experiments with surface acoustic waves at $\nu=1/2$. \cite{willett}
 
Although, when confronted with experiments, the composite fermion concept may be regarded as a huge success, the approach of Refs. \cite{lopez} - \cite{halperin} is not what one may call a {\it  theory} for the following reasons.\\

1) The appearance of the "Chern-Simons"  magnetic field   $B_{0} =  - 2mn\Phi_{0}$, which is independent of the real field $B$ and the strength of electron-electron interaction, is not justified.  (Why not do the same trick at $B=0$, or for non-interacting particles?)\\

2) The energy spectrum in the mean field approximation is obviously completely wrong, see Fig. 3. For example, at $\nu=2/5$ this approximation predicts energy gaps equal to $\hbar\omega_{c}/5$, which, in the limit of strong magnetic field, may differ from the true gaps ($\sim e^{2}/l$) as much as one chooses.  The first approximation being extraordinary far from the final result, its validity is very doubtful.\\ 

\begin{figure}

\hskip 25pt \epsfxsize=400pt{\epsffile{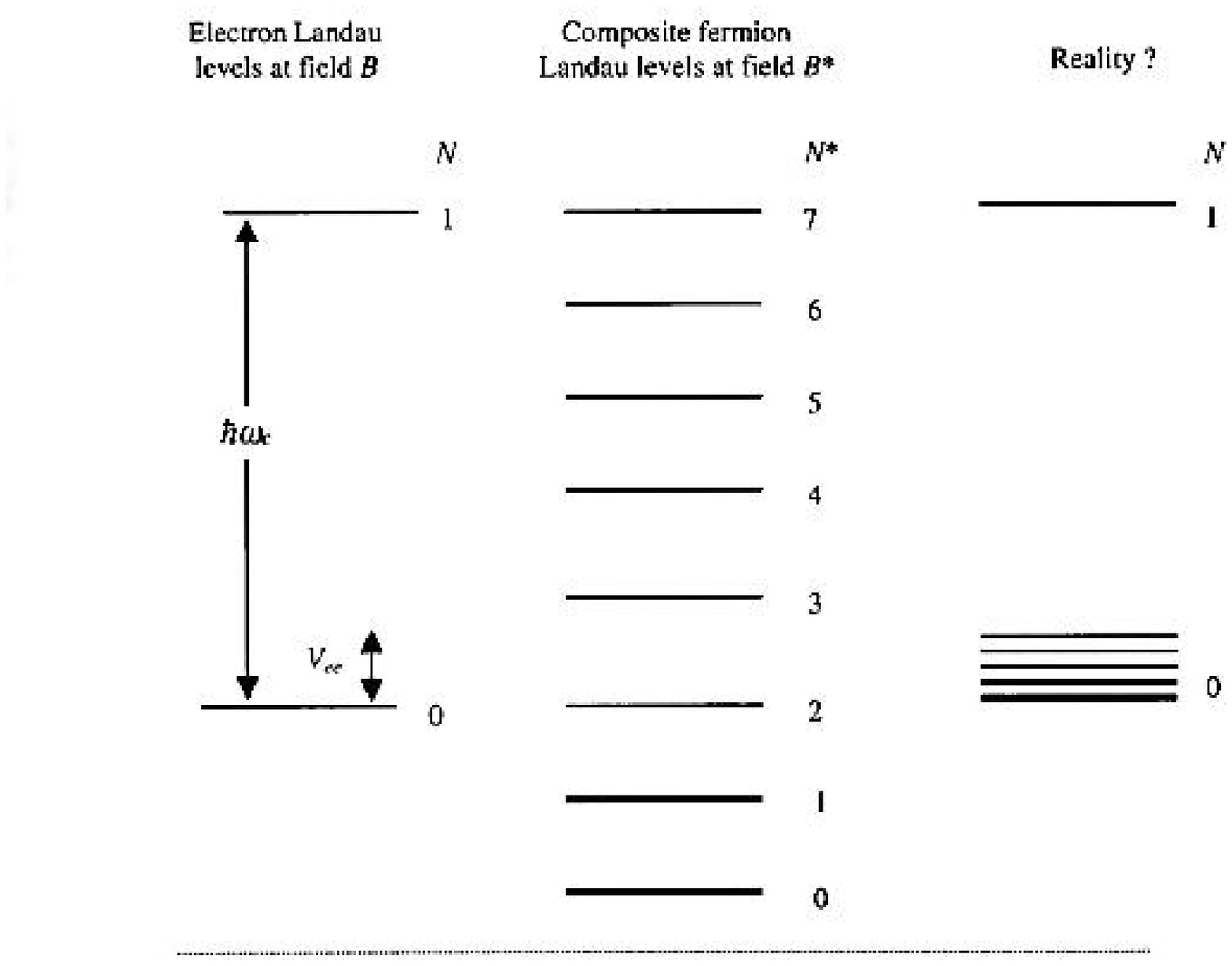}}
 
\vskip 15pt
 
\caption{The electron energy spectrum ($a$) and the composite fermion spectrum in the mean field approximation at $\nu=2/5$ ($b$). The electron interaction scale, $V_{ee}$, may be arbitrary small compared to the Landau  spacing . The reality presumably corresponds to ($c$).  Heavy lines represent occupied levels.}
 
\end{figure}

3) The effects of "gauge field fluctuations" and electron interactions cannot possibly cancel each other because they belong to quite different energy scales.  The mean field approximation consists in Eq. (9).  If one picks up the entire perturbation series (where the perturbation is the difference between $ns$ and the integer ${\cal N}$) one will return from the right-hand side of Eq. (9) to its left-hand side, which is equal to 1, thus annihilating the magnetic field $B_{0}$ that was arbitrary introduced.  However, if one makes some "approximation", one may arrive at {\it any} result depending on the (absolutely unjustified) choice of the terms to be neglected in the perturbation series.  Thus attempts to calculate the renormalized mass of the composite fermion are, in my opinion, meaningless.\\
  
4) The relation of the Chern-Simons approach to the Laughlin-Jain wavefunctions, which have the real advantage of being verified numerically, is unclear.  In the mean field approximation the electron wavefunction contains states in all Landau levels, which (in the strong field limit) does not have any physical sense, since higher levels are irrelevant.\\

Most of these objections are well known (see for example Ref. \cite{quinn}), and some attempts were made to derive an improved version of this theory, which involves states only in the lowest Landau level (see Refs. \cite{shankar}, \cite{read}), however, in my opinion, nothing simple and/or beautiful has yet emerged.   It is often stated that the Chern-Simons approach, after the renormalization of the composite fermion mass, correctly describes the low-energy part of the spectrum. What this really means (although nobody says so) is the picture in Fig. 3$c$: the {\it splitting} of the lowest Landau level due to electron interactions.   Indeed, if at $\nu=p/q$ with $q$ odd, the Landau level would split (in some sense, yet to be defined) into $q$ sublevels with degeneracy $M/q$ each, this could explain many experimental facts, which are currently accounted for by the composite fermion concept.  So far, nobody has shown this.  Such a splitting, if it exists, should be a consequence of some exact symmetry, rather than a result of arbitrary "approximations".\\

 \vskip 0.6cm                                                                 

{\bf 6. Model without magnetic field}\\

There is a simple, but theoretically important, question: are peculiarities at rational filling factors specific for the FQHE, or will they exist in other situations, when one has $M$ degenerate quantum states partially filled by $N$ interacting fermions (with repulsive interaction)?  Many such models can be proposed, but probably the simplest one is the following one-dimensional problem.

Consider $M$ degenerate one-particle states on a circle:

$$\psi_{k}(\phi)={1 \over \sqrt{2\pi}}\exp(ik\phi), \hspace{0.2in} k=0, 1, 2... M-1. \eqno{(10)}$$		

There are $N<M$ spinless fermions, which repel each other via some pair potential $U(\phi_{i}-\phi_{j})$. Find energy spectrum for a given $\nu$.  Although a one-particle Hamiltonian, having these degenerate eigenfunctions, may be easily presented, certainly the problem is somewhat artificial, and of course no Hall effect will exist.  However, the question we are interested in, is whether the energy spectrum for this system resembles that for the true FQHE system.  Will there be gaps in the excitation spectrum at $\nu=p/q$ with $q$ odd, gapless states at $\nu=1/2m$, and fractionally charged quasi-particles? 

It should be stressed that this model, designed as a caricature of the FQHE system, is very different from a model, in which fermions can occupy any of $M$ fixed sites on a circle. Although Wannier-type localized functions can be introduced by the relation
$$\Phi_{s}(\phi)=\Phi(\phi-\frac{2\pi s}{M})={1 \over \sqrt{M}}{\sum_{k=0}^{M-1}}\psi_{k}(\phi)\exp({-\frac{2\pi i}{ M}}ks), \hspace{0.2in}    s=0, 1, 2...M-1, \eqno{(11)}$$\\
the choice of the localization sites is arbitrary. (A different, but equivalent basic set can be obtained by shifting all the sites $\phi_{s}=2\pi s/M$ by an arbitrary angle $\alpha$.)

A crystal-like many-particle state may be constructed as a suitable determinant of these functions, which presumably will give the ground state for small enough $\nu$.  However, if $\nu=1/(2m+1)$ is not very small, a Laughlin-type function with uniform density should be preferable.  This function may be readily written as
$$\Psi_{1\over 2m+1}(\phi_{1},... \phi_{N})=A\prod_{1 \leq i<j \leq N} (\exp(i\phi_{i})-\exp(i\phi_{j}))^{2m+1},\hspace{0.1in} A^2= \frac {[(2m+1)!]^{N}}{(2\pi)^{N}((2m+1)N]!}.    \eqno{(12)}$$\\
Note that, in contrast to the case of the original Laughlin function, Eq. (3), for our simplified problem the normalization constant $A$ is found analytically, the corresponding normalization integral having been calculated by Dyson \cite{dyson} (see also Ref. \cite{mehta}). Exactly following Laughlin, quasi-holes and quasi-electrons may be introduced, and the same arguments will lead us to the conclusion that gaps in the excitation spectrum should appear for $\nu=1/(2m+1)$. Thus, it seems that, at least for the Laughlin states, there is no great difference between our model and the true FQHE system.

Interestingly, within our model an answer can be given to the question concerning the complimentary wavefunction at $\nu=1-1/(2m+1)$, which was discussed in Section 3.  It may be proved \cite{dyakonov} that, given the Laughlin-type function for $(2m+1)N=M$, Eq. (12), the complimentary wavefunction for $M-N$ particles at $\nu=2m/(2m+1)$, derived as indicated in Section 3, has a {\it similar} form:

$$\Psi_{2m\over 2m+1}(\phi_{1},... \phi_{M-N})=B{\prod_{1 \leq i<j \leq M-N}} (\exp(i\phi_{i})-\exp(i\phi_{j}))^{2m+1}, \eqno{(13)}$$\\
with a known normalization constant $B$.  The rhs of Eq. (13) contains powers of each $\exp(i\phi_{i})$ up to $(2m+1)(M-N-1)=2m(M-1)-1$, which is greater than $M$.  {\it These powers should be taken modulo M}, and this is a non-trivial {\it exact} result reminiscent of Jain's projection procedure \cite{jain}.
 
Using an expansion with respect to the basic set in Eq. (11), this result may be rewritten in another form in which the modulo $M$ rule is applied automatically.  In this basis Eq. (12) becomes:

$$\Psi_{1\over 2m+1}(\phi_{1},... \phi_{N})=\sum_{(s)}C(s_{1},...s_{N})\Phi_{s_{1}}(\phi_{1})...\Phi_{s_{N}}(\phi_{N}), \eqno{(14)}$$\\
where the sum is over all $s_{i}$ from $0$ to $M-1$ and the coefficients $C$ are given by:

$$C(s_{1},...s_{N})=A{\prod_{1 \leq i<j \leq N}}(\omega^{s_{i}}-\omega^{s_{j}})^{2m+1};\hspace{0.1in} \omega=\exp(\frac {2\pi i}{M}),\hspace{0.2in} A^{2}= \frac {[(2m+1)!]^{N}}{M^{N}[(2m+1)N]!}.    \eqno{(15)}$$\\
Then it can be proven, that the complimentary wavefunction, 

$$\Psi_{2m\over 2m+1}(\phi_{1},... \phi_{M-N})=\sum_{(s)}D(s_{1},...s_{M-N})\Phi_{s_{1}}(\phi_{1})...\Phi_{s_{M-N}}(\phi_{M-N}), \eqno{(16)}$$\\
has coefficients $D$, which have an appearance quite similar to Eq. (15):

$$D(s_{1},...s_{M-N})=B{\prod_{1 \leq i<j \leq M-N}}(\omega^{s_{i}}-\omega^{s_{j}})^{2m+1}. \eqno{(17)}$$\\
In this expression powers of $\omega^{s_{i}}$ higher than $M-1$ are automatically reduced to the interval $[0, M-1]$. 

It is esthetically pleasing that Eqs. (15) and (17) have exactly the same form, and this must have some profound reason which is not yet understood. One is tempted to extrapolate this result for other odd-denominator fillings and to suggest expressions like Eqs. (14), (15) as the ground state wavefunctions for all $\nu=p/q$ with $q=2m+1$. It can be proven \cite{dyakonov} that this conjecture is self-consistent, in the sense that {\it if} it is true for some $\nu$, it is also true for filling $1-\nu$.

It will be of considerable interest if somebody studies this simple model numerically.  If, as I believe, the results for odd and even denominator fractions will be quite similar to those for interacting electrons in the lowest Landau level, it could shed some light on the true FQHE problem.
 
What about composite fermions?  In my one-dimensional model there is no magnetic field, no fluxes to "attach" to electrons, and no loops to carry electrons around.  Clearly, for this model the language conventionally adopted by the FQHE theorists, as well as the way of thinking, should be strongly modified, and probably the same is true for the real FQHE problem.  It remains to be seen, if it is possible to introduce in a convincing way some quasi-particles, for which the initially $M$-fold degenerate level is split into $q$ sublevels.\\

\vskip 0.6cm

{\bf 7. Conclusions}\\

I have tried to convince the reader that what may be called the theory of the Fractional Quantum Hall Effect does not yet exist, although a large progress has been made in this enormously difficult theoretical problem due to the outstanding work in Refs. \cite{laughlin}, \cite{haldane} - \cite{halperin}, and many others, which were not discussed here. 
 
Currently, we are in an awkward position: on the one hand many experimental facts support Jain's  idea of composite fermions moving in a reduced effective magnetic field, and this is the only physical description available.  It certainly corresponds to some reality. On the other hand, nobody has really shown theoretically, apart from what may be described as wishful thinking, the existence of composite fermions, as (quasi) free particles. Moreover, this concept does not provide answers to a number of simple and fundamental questions; it does not even explain what is a composite fermion.

More efforts are needed to understand the underlying reality.  In my opinion, the properties of the energy spectrum at rational filling factors, which are responsible for the FQHE, are not specific for 2D electrons in a strong magnetic field, but should exist in many models with $N$ fermions occupying $M>N$ initially degenerate states, if the interaction is repulsive.  Studying such models, one of which was presented in Section 6, may help.

The true theory of the FQHE is yet to come, hopefully within the next 20 years.\\


%


\begin{references}


\bibitem{klitzing} K. von Klitzing, G. Dorda, and M. Pepper, Phys. Rev. Lett. {\bf 45}, 494 (1980).

\bibitem{tsui} D.C. Tsui, H.L. Stormer, and A.C. Gossard, Phys. Rev. Lett. {\bf 48}, 1559 (1982).

\bibitem{yoshioka} D. Yoshioka and P.A. Lee, Phys. Rev. B {\bf 27}, 4986 (1983).

\bibitem{laughlin} R. Laughlin, Phys. Rev. Lett. {\bf 50}, 1395 (1983).

\bibitem{su} W.P. Su and J.R. Schrieffer, Phys. Rev. Lett. {\bf 46}, 738 (1981).

\bibitem{stormer} H.L. Stormer, Physica B {\bf 177}, 401 (1992).

\bibitem{du} R.R. Du, H.L. Stormer, D.C. Tsui, L.N. Pfeiffer, and K.W. West, Phys. Rev. Lett. {\bf 70}, 2944 (1993).

\bibitem{haldane} F.D.M. Haldane, Phys. Rev. Lett. {\bf 51}, 605 (1983).

\bibitem{jain} J.K. Jain, Phys. Rev. Lett. {\bf 63}, 199 (1989); Phys. Rev. B {\bf 40}, 8079 (1989); Phys. Rev. B {\bf 41}, 7653 (1990).

\bibitem{lopez} A. Lopez and E. Fradkin, Phys. Rev. B {\bf 44}, 5246 (1991); {\it ibid} {\bf 47}, 7080 (1993).

\bibitem{zhang} V. Kalmeyer and S.-C. Zhang, Phys. Rev. B {\bf 46}, 9889 (1992).

\bibitem{halperin} B.I.Halperin, P.A. Lee, and N. Read, Phys. Rev. B {\bf 47}, 7312 (1993).

\bibitem{willett} R.L. Willett, Adv. Phys. {\bf 46}, 447 (1997).

\bibitem{fermions} {\it Composite Fermions}, Edited by O. Heinonen, World Scientific, Singapore (1998).

\bibitem{simon} S.H. Simon, {\it The Chern-Simons Fermi Liquid Description of Fractional Quantum Hall states}, {\it ibid}; LANL e-print cond-mat/9812186, (1998).

\bibitem{shankar} R. Shankar, {\it Theories of the Fractional Quantum Hall Effect}, LANL e-print cond-mat/0108271 (2001).

\bibitem{note1} Extensive numerical calculations, showing an excellent overlap (like 0.9999) of the Laughlin function with the $\nu=1/3, 1/5$ ground states obtained by exact diagonalization, convince us that the Laughlin function is the exact ground state at $N \rightarrow \infty$ for a wide class of repulsive potentials.  However nobody has been able to show this analytically, except for a special short-range potential, when it gives the exact ground state for arbitrary $N$.

\bibitem{note2} The relation between ground-state energies per particle, $E_{\nu}$ and $E_{1-\nu}$, reads: $\nu E_{\nu} - (1-\nu)E_{1-\nu} = (2\nu-1)E_{1}$, where $E_{1}$  is the energy per particle in a completely filled Landau level.  An exact formula for the chemical potential in a half-filled Landau level follows: $\mu_{1/2}=E_{1}$.  Apparently, these relations belong to the folklore. I was not able to find out, who has derived them for the first time.

\bibitem{note3} There is something odd about numerical calculations. To describe the {\it same} FQHE states, many {\it different} functions were proposed, which invariably showed excellent overlap with ground states found numerically.  For example, for $\nu=2/5$, various trial wavefunctions were proposed and checked numerically in Refs. \cite{morf} - \cite{landman}.  The reported overlap is always around or better than 0.99.  The reason for this universal success is not clear.  Does it mean that all these functions are, in fact, identical?  It should be noted that a polynomial of many variables might be written in many equivalent forms.

\bibitem{morf} R. Morf, N. d'Ambrumenil, and B.I. Halperin, Phys. Rev. B  {\bf 34}, 3037 (1986).

\bibitem{ambrum} N. d'Ambrumenil and R. Morf, Phys. Rev. B {\bf 40}, 6108 (1989).

\bibitem{girvin} D. Yoshioka, A.H. MacDonald, and S.M. Girvin, Phys. Rev. B {\bf 38}, 3636 (1988).

\bibitem{ginocchio} J.N. Ginocchio and W.C. Haxton, Phys. Rev. Lett. {\bf 77},  1568 (1996).

\bibitem{landman} C. Yannouleas and U. Landman, LANL e-print cond-mat/0202062 (2002).

\bibitem{kamilla} J.K. Jain and R.K. Kamilla, Phys. Rev. B {\bf 55}, 4895 (1997).

\bibitem{quinn} J.J. Quinn and A. Wojs, J. Phys.: Condens. Matter {\bf 12}, R265 (2000).

\bibitem{read} N. Read, Phys. Rev. B {\bf 58}, 16262 (1998).

\bibitem{dyson} F.J. Dyson, J. Math. Phys. {\bf 3}, 140 (1962).

\bibitem{mehta} M.L. Mehta, {\it Random Matrices}, Academic press (1991).

\bibitem{dyakonov} M.I. Dyakonov, to be published.

\end{references}
\end{document}